# Phonon abundance-stiffness-lifetime transition from the mode of heavy water to its confinement and hydration

Chang Q Sun[1*]

[1] EBEAM, Yangtze Normal University, Chongqing 408100, China (20161042@yznu.cn; pengyuan1030@sina.com); NOVITAS, Nanyang Technological University, Singapore 639798 (ecqsun@ntu.edu.sg)

**Abstract**

A combination of the temporally- and spatially-resolved phonon spectroscopy has enabled calibration of hydrogen bond transition from the vibration mode of heavy water to the core-shelled nanodroplet and the sub-nanosized ionic hydration shell in terms of phonon abundance-lifetime-stiffness. It is uncovered that charge injection by salt solvation and skin formation by molecular undercoordination (often called confinement) share the same supersolidity of H–O (D–O as a probe) bond contraction, O:H elongation, and electron polarization. The bond transition stems the solution viscosity, surface stress, and slows down the molecular dynamics. The skin reflection further hinders phonon energy dissipation and thus lengthens the phonon lifetime of the nanodroplet.

Contents

Content entry

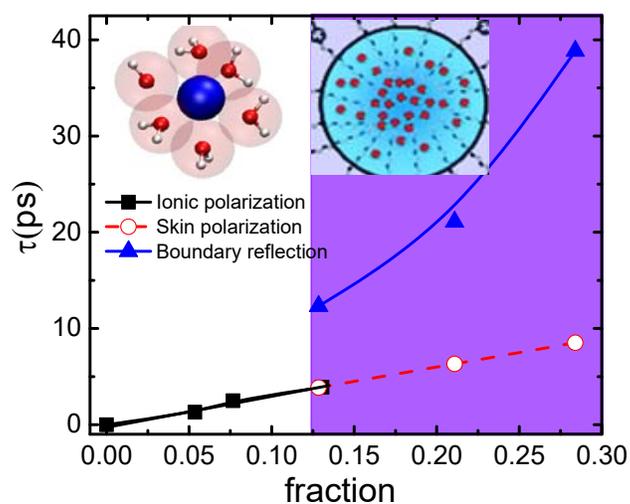

1. Introduction

Molecular performance at the solute-solvent interface and in the nanoconfined structures are ubiquitously important to subject areas of basic research and industrial applications such as health care, food, medication, nutrition, anti-pollution, anti-corrosion and friction [1-5]. However, high-resolution detection and consistent insight into the fascinations stay challenge. Water molecules subjecting to undercoordination in clusters [6, 7], droplets [8], nanocavities [9], and nanotubes [10], cloud/gel surface [11] and those stayed surrounding ionic solutes of salts [12] form a phase that is unusual to the bulk [13]. Recent evidence [13] shows that two forms of 2D liquid water can arise in deeply supercooled states, but separated by the solid phase of bilayer amorphous ice whose melting line exhibits the isochore end point. The phase is featured by its higher viscosity, lower diffusivity, slower molecular motion dynamics, shorter H–O or D–O bond but longer phonon lifetime because of the ionic and skin polarization [14].

Experimental detection [15] and molecular dynamics (MD) computations [16] showed consistently that both salt solvation and water confinement not only slow down MD dynamics characterized by the phonon lifetime but also shift the H–O or the D–O phonon frequency to higher frequencies. One needs to answer why do salt solvation and nanoconfinement transit the phonon lifetime and phonon stiffness in the same manner and what intrinsically dictates the chemical and physical properties of the confined and the hydrating water in solution.

According to Pauling [17], the nature of the chemical bond bridges the structure and property of a crystal and molecule. Therefore, bond formation and relaxation and the associated energetics, localization, entrapment, and polarization of electrons mediate the macroscopic performance of substance accordingly [18]. The recent hydrogen bond (O:H–O) cooperativity notion [14, 19-21] has enabled a resolution to multiple anomalies of water and ice. A combination of the Lagrangian mechanics, MD and DFT computations with the static phonon spectrometrics has enabled quantification of O:H–O transition from the mode of ordinary water to the conditioned states. Obtained information includes the fraction, stiffness, and fluctuation order transition upon perturbation and their consequence on the solution viscosity, surface stress, phase boundary dispersity, and the critical pressure and temperature for phase transition [19]. One must focus on the bond relaxation and electron polarization in the skin region or in the hydration volume as degrees of independent freedom.

We show in this communication the strategies and findings gained by interplaying the time–resolved infrared spectroscopy and the zone-resolved differential phonon spectroscopy (z-DPS) [12] toward the D–O phonon abundance–lifetime–stiffness cooperative transition upon NaBr solvation and confined in reverted micelles, with extending the situation focused on the phonon relaxation time from the perspective of molecular dynamics [15].

2. Hydrogen bond cooperative relaxation

2.1. Hydrogen bond cooperativity

As a strongly correlated and fluctuating system, water prefers the statistic mean of the tetrahedrally-coordinated, two-phase structure in a bulk-skin or a core-shell manner of the same molecular geometry but different O:H–O bond segmental lengths [14, 19]. The O:H–O bond integrates the intermolecular weaker O:H nonbond (or called van der Waals bond with ~0.1 eV energy) and the intramolecular stronger H–O polar-covalent bond (~4.0 eV) with asymmetrical, short-range interactions and coupled by the Coulomb repulsion between electron pairs on adjacent oxygen ions [14], see Figure 1a inset the elongated O:H–O bond by ionic polarization [12] and molecular undercoordination [8].

The O:H nonbond and the H–O bond segmental disparity and the O–O electrostatic correlation allow the segmented O:H–O bond to relax oppositely – an external stimulus dislocates both O ions in the same direction but by different amounts. The softer O:H nonbond always relaxes more than the stiffer H–O bond with respect to the $H^+$ coordination origin. The O:H–O bond cooperativity determines properties of water and ice under external stimulus such as molecular undercoordination [4, 22-25], mechanical compression [26-30], thermal excitation [31-33], solvation [34, 35] and determines the molecular behavior such as solute and water molecular thermal fluctuation, phonon relaxation, solute drift motion dynamics, solution viscosity and surface stress.

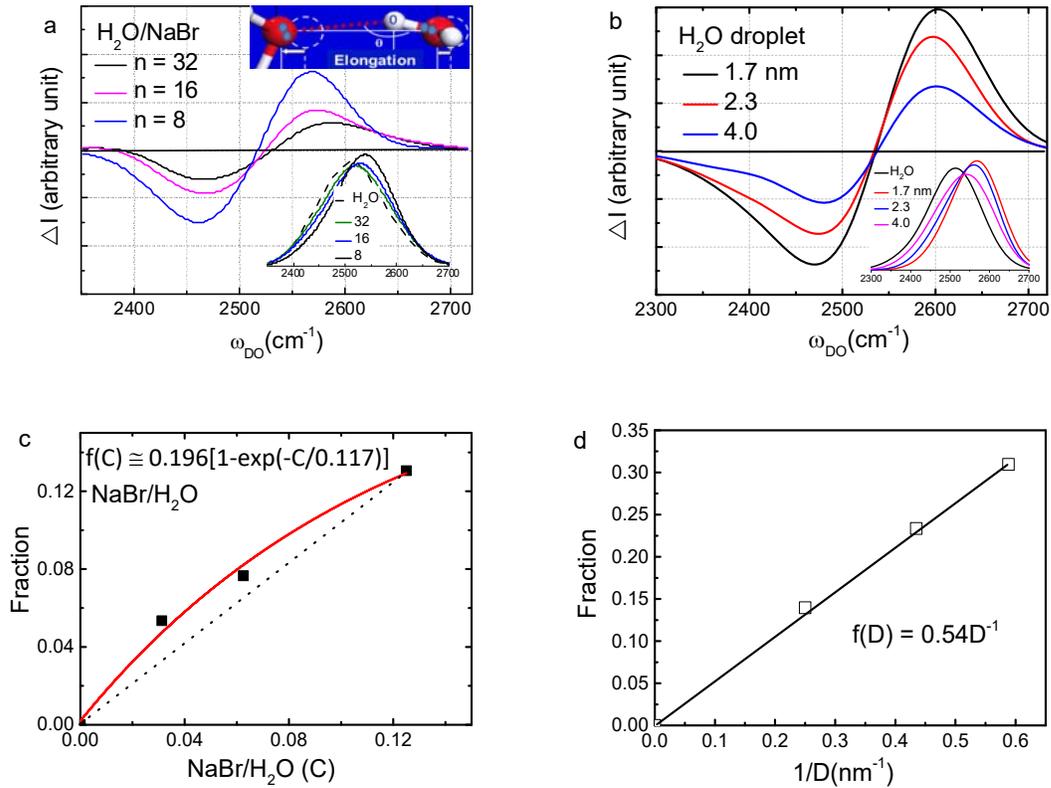

Figure 1. The D–O phonon (a, b) z-DPS and (c, d) the fraction coefficient for (a, c) the concentrated NaBr solutions and (b, d) the sized water droplets. Insets a and b show the D–O peaks presented in [15]. Inset a illustrates the O:H–O bond elongation by polarization due to salt solvation and molecular undercoordination [8, 12].

## 2.2. Solvation and confinement bonding dynamics

Intensive investigation has confirmed that salt solvation and molecular undercoordination share the same effect on intramolecular covalent bond contraction, intermolecular nonbond elongation, and charge polarization, associated with high–frequency phonon blueshift and low-frequency phonon redshift. The polarization results in the supersolidity in the ionic solute hydration volume [12] and in the covering sheet of water ice [8, 9].

The concept of supersolidity was initially extended from the $^4$He fragment at mK temperatures, demonstrating elastic, repulsive and frictionless between the contact motion of $^4$He segments [36] because of atomic undercoordination induced local densification of charge and energy and the associated polarization [37]. The concepts of supersolidity and quasisolidity were firstly defined for water and ice in 2013 by Sun et al [8, 32] and then be continually verified.

The quasisolidity describes phase transition from Liquid density maximum of one gcm$^{-3}$ at 4 °C to Solid density minimum of 0.92 gcm$^{-3}$ at –15 °C [32]. The quasisolid phase demonstrates the cooling expansion because of the segmental specific heat ratio $\eta_L/\eta_H < 1$, the H–O bond contraction drives the O:H expansion. The $\eta_L$ and $\eta_H$ are the specific heat for the O:H and the H–O part, respectively. At a certain temperature, the segment having lower specific heat follows the regular thermal expansion and the other part relaxes oppositely because of the O–O coupling in the O:H–O bond.

The supersolidity features the behavior of water and ice under polarization by coordination number reduction or charge injection. When the nearest CN number is less than four the H–O bond contracts spontaneously associated with O:H elongation and strong polarization. At the surface, the H–O bond contracts from 1.00 to 0.95 Å and the O:H expands from 1.70 to 1.95 Å associated with the O:H vibration frequency transiting from 200 to 75 cm$^{-1}$ and the H–O from 3200 to 3450 cm$^{-1}$ [38]. The shortened H–O bond shifts its vibration frequency to a higher value that increases again with further reduction of the molecular CN.

A salt molecule dissolves into an anion and a cation distributed regularly in the yet seemingly disordered liquid with or without skin preferential occupation. Ions serve each as source of electric field that aligns, clusters, stretches, and polarizes their neighboring solvent molecules to form the supersolid [8] or semirigid [39, 40] hydration shells with H–O contraction and O:H elongation. The polarized solvent molecules in the hydration shells screen in turn the solute electric field.

The O:H–O bond between undercoordinated water molecules in the skin of bulk water and ice, nanobubbles, or nanodroplet is subject to H–O contraction and O:H elongation and strong polarization, which triggers the slipperiness of ice and the strength and stability of nanobubbles and nanodroplets [8, 9]. For the skin of water and ice, the H–O phonon shifts from 3200/3150 to 3450 cm$^{-1}$. The frequency shifts from 3200 to 3650 cm$^{-1}$ when the N of the $(H_2O)_N$ cluster drops from 6 to 1 [41-43]. The skin polarization depresses the bound energy of nonbonding electrons from the bulk value of 3.2 to the skin of 1.6 eV [14]. The bound energy drops further as the droplet size is reduced.

3. Dynamic and static phonon spectroscopies

The pump–probe time–dependent phonon spectroscopy probes the decay time of a known intramolecular (H–O or D–O) vibration phonon population to derive the molecular motion dynamics through the solution viscosity and Stokes–Einstein relation for the molecular drift diffusivity [44], which is very much the same to optical fluorescent spectroscopy [45]. The signal lifetime is proportional in a way to the density and distribution of the defects and impurities. The impurity or defect states prevent the thermalization of the electrons transiting from the excited states to the ground for exciton (or electron–hole pair) recombination. One switches off/on the pump/probe simultaneously and monitors the population decay that features the rate of vibration energy dissipation during the wave propagation in the solution.

The z-DPS is obtained by subtracting the D–O peak from those of solutions and those of water confined in nanopores upon all of them being spectral areal normalized [12]. This subtraction resolves the fraction of bonds transiting from the mode of ordinary water to the states of polarization by confinement [8] or solvation [12]. Therefore, combining the static z-DPS and the dynamic ultrafast phonon spectrometrics could correlate the molecular spatial and temporal performance to the O:H–O bond relaxation and transition upon salt solvation and nanoscale confinement with detailed information on the hydrating and the confined bonding and electronic states that inhibit phonon propagation.

4. Phonon abundance-stiffness transition

One often measures the phonon spectroscopy of a $D_2O$ and $H_2O$ mixture for better resolution. Ultrafast

pump–probe IR spectroscopy detection of Park et al [15] revealed that 10% NaBr solvation slows down the hydrogen bond network relaxation in the 5%D$_2$O + 95%H$_2$O solvent by a factor of 3 associated with D–O bond stiffness transition from 2480 to 2560~2580 cm$^{-1}$. Contrastingly, water molecules confined in the reverse micelles relax even slower than that in the ionic hydration volume associated with blue shift of the D–O phonons from 2480 to 2600 cm$^{-1}$, see Figure 1a and b insets. Observations [15] confirmed that water molecular motion dynamics in the 4 nm diameter nanopools of the neutral and the ionic reverse micelles is almost identical, which confirms that confinement by the cell–water interface is a primary factor governing the dynamics of nanoscopic water rather than the presence of charged groups at the interface. The interface plays the dominant yet unclear role in determining the hydrogen bonding dynamics, whereas the chemical nature of the interface plays a secondary role [15].

The z-DPS profiles for the NaBr solutions and water nanodroplets shown in Figure 1a and b revealed that the structure order characterized by the z-DPS linewidth and the transition of phonon abundance increase with the NaBr concentration and droplet size reduction. Inset a illustrates the O:H–O elongation under ionic polarization and molecular undercoordination, proceeded by lengthening the O:H and shortening the H–O [8, 12]. Both the z-DPS measurements and density functional theory calculations show consistently the skin density is 0.75 gcm$^{-3}$ [38] for ice and water, which agrees with the findings documented in [13].

The phonon peak frequency follows the $\omega^2 \propto E(\mu d^2)^{-1}$ relation with $\mu$ being reduced mass and $E$ the binding energy and $d$ the length of the H–O or D–O vibrating dimer. Therefore, the phonon abundance transition from the valley to the peak in both cases show that the D–O bond turns to be shorter and stiffer in the hydration shell and in the droplet skin.

An integration of the z-DPS peak gives rise to the fraction of bonds transiting from the mode of water to polarization in the ionic hydration volume and in the skin of the droplets, as shown in Figure 1c and d. The fraction coefficients follow the relations:

$$f_{NaBr}(C) \propto 0.196[1-\exp(-C/0.117)]$$

$$f_{droplet}(D) = 0.54D^{-1}$$

Excitingly, the $f_{NaBr}(C)$ follows exactly the general form for salt solutions [12] and the $f_{droplet}(D)$ follows the universal shell/core volume ratio for nanostructures [37].

It has been shown in [12] that the $f_{NaBr}(C) = f_{Na}(C) + f_{Br}(C)$ results from the invariant $Na^+$ hydration volume and the variant $Br^-$ hydration volume. Solvation dissolves the NaBr into $Na^+$ and $Br^-$ ions that serve each as a center of electric field that aligns, clusters, stretches and polarizes its surrounding O:H–O bonds to form the supersolid hydration shells. The slope of $f_x(C)$ is in proportional to the hydration volume or the local hydration electric field. The small $Na^+$ solute forms a constantly sized hydration droplet without responding to interference of other ions because its hydrating $H_2O$ dipoles fully screen its electric field. The linear $f_{Na}(C)$ contributes to the linear part of the Jones–Dole viscosity [12, 46].

However, the number inadequacy of the highly-ordered hydration $H_2O$ dipoles partially screens the large $Br^-$. The $Br^-$ then interacts repulsively one another, which weakens its electric field and the $f_{Br}(C)$ approaches saturation at higher solute concentration. Therefore, the $Br^- - Br^-$ repulsion weakens the local electric field of the $Br^-$ anion, which softens the z-DPS phonon from 2580 to 2560 $cm^{-1}$ at higher solute concentrations [47], and gives rise to the exponential $f_{NaBr}(C)$ toward saturation, in Figure 1c.

Molecular undercoordination shares the same effect of electronic polarization. The $f_{droplet}(D) = 0.54D^{-1}$ indicates that the skin H–O bond contraction dictates the nanodroplet performance. The $f_{droplet}(D)$ clarifies the presence of the supersolid covering sheet of a constant thickness $\Delta R$. From Figure 1d, one can estimate the shell thickness $\Delta R$ of a spherical droplet of $V \propto R^3$: $f(R) \propto \Delta V/V = \Delta N/N = 3\Delta R/R$, and the skin thickness $\Delta R_{skin} = f(R)R/3 = 0.90$ Å is exactly the dangling H–O bond length [14] featured at 3610 $cm^{-1}$ wavenumber of vibration [8]. The next shortest H–O bond in the skin is about 0.95 Å featured at 3550 $cm^{-1}$ [48]. The z-DPS distills only the first hydration shell and the outermost layer of a surface [49] without discriminating the intermediate region between the bulk and the outermost layer.

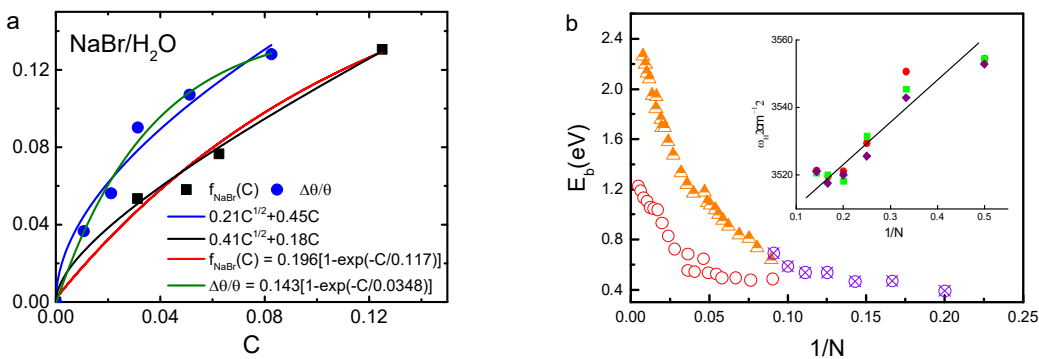

Figure 2. (a) NaBr concentration resolved contact angle $\Delta\theta(C)/\theta(0)$ (surface stress) [50] and the $f_{NaBr}(C)$ match the viscosity of Jones–Dole premise $\Delta\eta(C)/\eta(0)$ [51]. (b) $(H_2O)_N$ size resolved shift of the nonbonding electron bound energy [52-54] and the H–O vibration frequency [41, 55-57].

Figure 2a shows that the relative contact angle (surface stress, $\Delta\theta(C)/\theta(0)$) and the $f_{NaBr}(C)$ follow the relative viscosity of Jones–Dole, $\Delta\eta(C)/\eta(0) = AC^{1/2} + BC$ [51], with terms of $C^{1/2}$ and C arising from polarization by the larger anions and smaller cations, respectively [12]. The $(H_2O)_N$ size resolved H–O vibration frequency and nonbonding electron bound energy, in Figure 2b show the expected trends of change. Therefore, the skin molecular undercoordination and salt solute-solvent interface share the same supersolidity due to H–O bond contraction and electron polarization, albeit their extent.

5. Polarization and skin reflection hinder the dynamics

Ultrafast IR spectroscopy revealed that the molecular orientation relaxation (residing) time increases from 2.6 to 3.9 and 6.7 ps as the water transit into NaBr solution with concentration increasing from 32 to 8 $H_2O$ per NaBr and increases from 18 to 50 ps when the water droplet size is reduced from 4.0 to 1.7 nm [15]. Table 1 lists the relaxation time $\tau$ and the fraction coefficients $f(x = D$ or $C)$ for bond transition from the mode of ordinary water to the supersolid states in the solute-solvent interfaces and in the covering sheet of the droplets.

Likewise, a proton mobility investigation [58] revealed that proton diffusivity slows down significantly with decreasing droplet size, see Figure 3a, being in the form of $D^+_H = 1.585D$, which means that the diffusivity varies linearly with droplet size. For water droplet smaller than 1 nm, the diffusivity is about two orders lower than it is in bulk water. The more rigid hydrogen–bond network of the confined water or the high viscosity of the droplet is suggested to lower the proton mobility. Furthermore, it is harder to break the even shorter and stronger H–O bond for the undercoordinated water molecules [14].

Figure 3b shows the linear $\tau(x) \propto f(x)$ correspondence with a slope of $d\tau/df \cong 30$ for the NaBr solution, which clarifies that the phonon lifetime depends linearly on the number fraction of O:H–O bonds transiting from the mode of water to polarization. The supersolidity of the hydration hinders phonon propagation or vibration energy dissipation and thus elongates its lifetime in the pump–probe detection process, being in principle the same to the optical fluorescent spectroscopy [45]. This understanding may deepen the insight into the mechanism of molecular residing in the supersolid hydrating and confined volume.

The striking difference between a water droplet and a salt solution is the correlation between the phonon lifetime $\tau$ and the fraction coefficient. The slope $d\tau(f)/df \approx 30$ remains constant for the NaBr solutions, but the $d\tau(f)/df$ is different substantially for the water droplet. The $\tau$ values for the droplets are much greater than that for the NaBr solutions though both salt solvation and molecular undercoordination derive the same supersolid phase [8, 12].

Besides the effect of polarization, the phonon spatial confinement by the droplet skin boundary shall inhibit the phonon propagation. The outwardly propagating phonon wave partially transmitted and partially reflected by the boundary. The superposition of the propagating and reflecting waves form then the weakly time dependent standing wave with a longer lifetime. Ideally, the lifetime of a standing wave is infinity. The vibration energy dissipates slower than it does in the open salt solutions because of the boundary reflection.

An extrapolation of the $\tau(f)$ due to polarization to the droplet regime leads to the excessive $\tau(f)_{droplet} -$

$f_{droplet}d\tau(f)/df_{solution} = \tau(f)_{confinement}$, which shall discriminate the geometric confinement of the droplet boundary from skin polarization, as illustrated in Figure 3b. Therefore, the population decay of D–O phonons in the nanodroplet consists of two components. One is the skin supersolidity and the other the geometric reflection.

Table 1. Comparison of the D–O bond orientation relaxation time and the fraction of bonds transferred from the mode of pure water to the polarized states in NaBr solutions and water nanodroplets. For solution, $d\tau/df \approx 30$ arises from the polarization. The excessive $\Delta\tau_{confinement} = \tau_{droplet} - \tau_{bulk} - f_{droplet}(d\tau/df)_{solution}$ arises from the geometric phonon confinement of the by the droplet grain boundaries that prohibiting phonon propagation.

| sample | | $\tau$(ps) [15] | $\omega_M$ (z-DPS) | f(x) | $\tau_{polarization}$ | $\tau_{confinement}$ | $\Delta R_{skin}$ |
|---|---|---|---|---|---|---|---|
| pure water | | 2.6 | 0 | – | 0 | – | – |
| H$_2$O/NaBr | 32 | 3.9 | 2586 | 0.0535 | 1.3 | – | – |
| | 16 | 5.1 | 2574 | 0.0766 | 2.5 | – | – |
| | 8 | 6.7 | 2568 | 0.1306 | 3.9 | | |
| Nanograin (nm) | 4.0 | 18 | 2600 | 0.1285 | 3.86 | 12.32 | 0.09 |
| | 2.3 | 30 | 2597 | 0.2108 | 6.32 | 21.08 | 0.09 |
| | 1.7 | 50 | 2602 | 0.2839 | 8.52 | 38.84 | 0.09 |

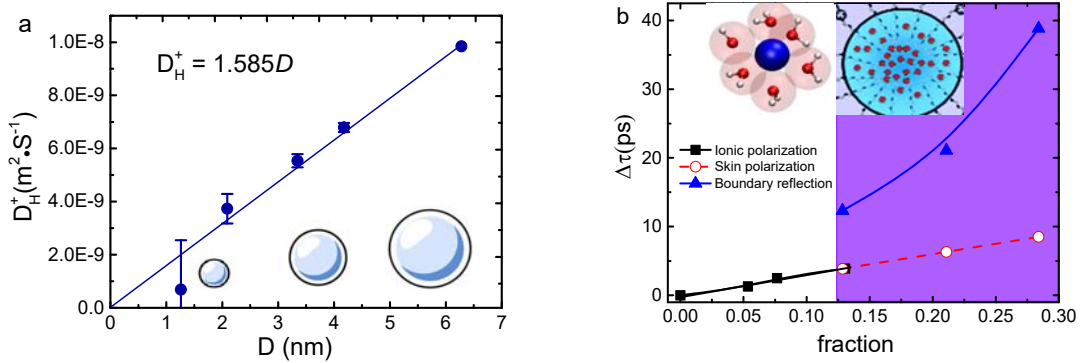

Figure 3. (a) Water droplet–size resolved D$^+$ diffusivity [58] and (d) the D–O phonon lifetime as a function of the fraction coefficient with discrimination of the supersolidity polarization due to solvation and molecular undercoordination and water droplet boundary reflection (triangles), $\Delta\tau_{solution}$ = $\tau_{solution} - \tau_{bulk}$ and $\Delta\tau_{confinement} = \tau_{droplet} - \tau_{bulk} - f_{droplet}(d\tau/df)_{solution}$. Insets in d illustrate the droplet confined in a reverse micelle [15] and the supersolid solute hydration shell.

6.   Perspective

A combination of the static DPS and the dynamic ultrafast phonon spectroscopy has reconciled, for the first time, the phonon dynamics in terms of hydrogen bond transition and molecular temporal and spatial performance, which results the hither to comprehensive and quantitative information on the phonon abundance–lifetime–stiffness transition from the ordinary water to the solute-solvent interface and to the skin of nanodroplet. Water molecular undercoordination and salt solvation share the same supersolidity of O:H–O bond cooperative relaxation and electron polarization.

The number fraction of bond transition to the solute-solvent interface $f_{NaBr}(C) \propto 1-\exp(-C/C_0)$ follows universal form for the concentrated salt solutions and the $f_{droplet}(D) = 3\Delta D/D$ follows the skin/volume ratio for the sized nanostructures with the skin thickness $\Delta D/2 = 0.90$ Å of the dangling H–O bond length. The H–O /D–O bond contraction stiffens their phonon frequencies. Polarization raises the solution viscosity and surface stress and slows down the molecular diffusivity and depressed binding energy of the nonbonding electrons.

Polarization hinders molecular dynamics and lengthens the phonon lifetime. Besides the supersolidity contribution, the skin reflection hinders vibration energy dissipation by forming the weakly time dependent standing waves. The phonon lifetime $\tau_{solution}(C) \propto f_{NaBr}(C)$ features the effect of ionic polarization and the $\tau_{droplet}(D) = \tau_{polarization}(D) + \tau_{confinement}(D)$ discriminates the skin polarization from the droplet boundary reflection.

Exercises and achievements show the promise and profoundness of amplifying the surface and interface from macroscopic to atomic, from solidus to the aqueous, and from static to dynamic by combining the ultrafast and the differential phonon spectrometrics. The combination of both spectroscopies offers information that is beyond the ability of either of them alone.

Acknowledgement

Financial support received from Natural Science Foundation (Nos. 11872052(YL); 21875024(CQ)), the Science Challenge Project (No. TZ2016001) of China is acknowledged.